# Fast Phase-manipulation of the Single Nuclear Spin in Solids by Rotating Fields

T. Shimo-Oka,[1] Y. Tokura,[2] Y. Suzuki,[3] N. Mizuochi[1]

[1]*Institute for Chemical Research, Kyoto University, Uji, Kyoto 610-0011, Japan*

[2]*Graduate School of Pure and Applied Sciences, University of Tsukuba, Tsukuba, Ibaraki 305-8571, Japan*

[3]*Graduate School of Engineering Science, Osaka University, Toyonaka, Osaka 560-8531, Japan*

**Abstract**

  We propose fast phase-gates of single nuclear spins interacting with single electron spins. The gate operation utilizes geometric phase shifts of the electron spin induced by fast/slow rotating fields; the path difference depending on nuclear spin states enables nuclear phase shifts. The gate time is inversely proportional to the frequency of the slow rotating field. As an example, we use nitrogen-vacancy centers in diamond, and show the phase-gate time orders of magnitude shorter than previously reported. We also show the robustness of the gate against decoherence and systematic errors.

## I. NTRODUCTION

The realization of long-lived qubits with sufficient operability is an essential issue in the field of quantum information. Nuclear spins are one of the most attractive platforms for their prominent spin coherence. On the other hand, nuclear Rabi oscillations are much slower than those of electron spins because of their small magnetic moment. As a solution to this problem, nuclear spin phase-controls via electron spin transition have been theoretically proposed [1] and experimentally demonstrated [2]. The phase gate is enabled by the hyperfine coupling and off-resonant transitions of the electron spin. Using this method, the gate time can be improved up to the inverse of the hyperfine constant.

Here, we propose a new method for fast phase control of single nuclear spins. As the gate time is not limited by the hyperfine constant, in principle this method offers gate operations that are orders of magnitude faster than previously reported. The proposed gate operation utilizes geometric phase shifts of single electron spins [3–6]; here, the electron spin dynamics are controlled by rotating fields. As the time evolution of the electron spin depends on nuclear spin states, the rotating field enables nuclear spin phase-gates. We also found that our gates are robust against decoherence and systematic errors. Finally, we show how to implement conditional phase gates using two different nuclear spins.

In the proposed method, we focus on single nuclear spins interacting with single electron spins of nitrogen-vacancy (NV) centers in diamond [7–10]. The NV electron spins are well known for their outstanding spin coherence ($T_2 > 1$ ms at room temperature (RT)) [11]. The electron spin couples to a wide range of forces—magnetic, optical, and electrical. Optical excitations allow for spin initialization and readout [12]. Coherent coupling with single photons [13] is used at low temperature in order to generate quantum network among separate NV centers [14,15]. Stark shifts in the excited state make spin-sublevel shifts in the ground state via spin–orbit couplings at RT [16], and the electric-field effect enables electron-spin controls between $m_s =1$ and $m_s = -1$ states [17]. In the proposed methods, the electron spin dynamics is controlled by oscillating electric/magnetic fields; for reasons that are discussed later, the electric/magnetic-field control is superior to solely magnetic-field controls for fast operations.

## II. SPIN HAMILTONIAN AND DYNAMICS

In the proposed gating method, an NV electron spin ($S = 1$) interacts with a single nitrogen-15 ($^{15}$N) nuclear spin ($I = 1/2$). The ground-state spin Hamiltonian is described [16,18,19] as ($\hbar =1$ from now on)

$$H_0 = DS_z^2 + \gamma_e BS_z + d_\perp E_x \left(S_x^2 - S_y^2\right) + d_\perp E_y \left(S_x S_y + S_y S_x\right) + A_\parallel S_z I_z + \frac{A_\perp}{2}\left(S_+ I_- + S_- I_+\right), \quad (1)$$

where $\gamma_e$ is the electron gyromagnetic ratio, $B$ is the magnetic field oriented along the NV axis. $d_\perp/2\pi$ ( $= 17$ Hz cm/V ) is the electric dipole moment, $E_{x,y}$ are electric fields with directions perpendicular to the NV axis. $D$ is the zero-field splitting parameter, and $A_\parallel$ ($A_\perp$) is the hyperfine coupling constant. These parameters are set to their reported values of $D/2\pi = 2.87$ GHz, $A_\parallel/2\pi = 3.03$ MHz, and $A_\perp/2\pi =3.65$ MHz. Here we omit the nuclear Zeeman terms for simplicity as these terms are much smaller than the others. In the following, we use $|\ >_E$ ($|\ >_N$) to represent electron (nitrogen nuclear) spin states. We also assume that, under the following conditions, $(A_\perp/(D-\gamma_e B))^2 \sim 10^{-6}$, the large energy splitting allows to neglect the electron-nuclear flip-flop terms of Eq. (1), and the electron spin can be described as a two-level system. The Hamiltonian can be thus simplified in the interaction picture of the zero-field splitting:

$$H = \gamma_e B\sigma_z + d_\perp E_x \sigma_x + d_\perp E_y \sigma_y + A_\parallel \sigma_z I_z, \quad (2)$$

where $\sigma_i$ ($i=x, y, z$) are the Pauli operators.

In the proposed method, spin input states are prepared under static electric/magnetic fields. During phase gating of the nuclear spin, we set the static field to zero and apply oscillating electric/magnetic fields $\gamma_e B = \omega_1 \cos(\omega t)$, $d_\perp E_x = \omega_1 \sin(\omega t)\cos(\varphi)$, and $d_\perp E_y = \omega_1 \sin(\omega t)\sin(\varphi)$. The oscillation of each field amplitude induces an inertial force, which controls NV spin states. In the laboratory frame, the Hamiltonian is then re-written as,

$$H(t) = \vec{\omega}_1(\omega t, \varphi) \cdot \vec{\sigma} + A_\parallel \sigma_z I_z, \qquad (3)$$

where $\vec{\omega}_1(\omega t, \varphi)$ represents the amplitude of each oscillating field on the polar coordinate system. The oscillation of the electric/magnetic field plays a key role in the proposed gate operation; if the amplitude $\omega_1$ is too small, $(\omega_1/A_\parallel)^2 \leq 10^{-7}$, the effect of the oscillating field is suppressed by the orthogonal static magnetic field (Appendix A); thus, we set the amplitude to satisfy the condition, $(\omega_1/A_\parallel)^2 \gg 10^{-7}$. On the other hand, if we use solely magnetic-field control instead of electric/magnetic-field control, the large zero-field splitting makes it difficult to observe the effect of transverse magnetic fields.

The three oscillating-field components of Eq. (3) correspond to one rotating field, whose rotational axis is perpendicular to $z$-axis of the NV-defect coordinate system, ($x$, $y$, $z$) (Fig. 1(a)). Thus, it is useful to set another coordinate system, ($X'$, $Y'$, $Z'$), where $Z'$ indicates the rotational axis. The coordinate transform is described as

$$\begin{pmatrix} \sigma_x \\ \sigma_y \\ \sigma_z \end{pmatrix} = \begin{pmatrix} \cos\varphi & -\sin\varphi & \\ \sin\varphi & \cos\varphi & \\ & & 1 \end{pmatrix} \begin{pmatrix} \sigma'_X(\varphi) \\ \sigma'_Y(\varphi) \\ \sigma'_Z(\varphi) \end{pmatrix}. \qquad (4)$$

The Hamiltonian in the ($X'$, $Y'$, $Z'$)-coordinate system is then represented as

$$H(t) = \omega_1 \left(\cos(\omega t)\sigma'_X(\varphi) + \sin(\omega t)\sigma'_Y(\varphi)\right) + A_\parallel \sigma'_X(\varphi) I_z. \qquad (5)$$

In analyzing the effect of this rotating field, it is useful to employ a rotating-frame. Using a unitary operator

$$U_1(t) = \exp\left(-i\frac{\omega t}{2}\sigma'_z(\varphi)\right), \tag{6}$$

the rotating-frame Hamiltonian is described as $U_1^\dagger(t)H(t)U_1(t) - iU_1^\dagger(t)\,d/dt\,U_1(t) = H_1 + V_1(t)$, where

$$H_1 = \omega_1\sigma'_X(\varphi) - \frac{\omega}{2}\sigma'_z(\varphi), \tag{7a}$$

$$V_1(t) = A_\parallel \exp\left(i\frac{\omega t}{2}\sigma'_z(\varphi)\right)\sigma'_X(\varphi)\exp\left(-i\frac{\omega t}{2}\sigma'_z(\varphi)\right)I_z. \tag{7b}$$

Here, $H_1$ are the non-perturbation terms, and $V_1(t)$ is the perturbation term. We next represent the operator $U_1(t)$ in the $(x, y, z)$-coordinate system. From Eq. (4), the matrix, $\sigma'_z(\varphi)$, is written as $\sigma'_z(\varphi) = \exp(-i\varphi\sigma_z/2)\,\sigma_y\,\exp(i\varphi\sigma_z/2)$. The operator $U_1(t)$ can then be described in the $(x, y, z)$-coordinate system as

$$U_1(t) = \exp\left(-i\frac{\varphi}{2}\sigma_z\right)\exp\left(-i\frac{\omega t}{2}\sigma_y\right)\exp\left(i\frac{\varphi}{2}\sigma_z\right). \tag{8}$$

The rotating frame Hamiltonian of Eq. (7) denotes that the electron spin interacts with the inertial force of the rotational field, $-iU_1^\dagger(t)\,d/dt\,U_1(t)$. However, in rotating fields with too high frequency, $|A_\parallel/\omega|, |\omega_1/\omega| \ll 1$, dynamic and geometric phases induced by the inertial force are nearly canceled [20]. Thus, it is difficult to perform fast phase-gates by one rotating field with too high frequency. By contrast, we next show that using two rotating fields with different frequencies, allows for accumulation of the geometric phase without cancellation, even in rotational fields with sufficiently high frequency.

We consider that when the rotating-frame Hamiltonian changes adiabatically, the spin follows it [20]. This inertial-force change can be implemented through phase shifts of the electric field, $\varphi \to \omega' t$, which satisfy the following condition, $|\omega'/\omega| \ll 1$. From Eq. (8), shifting the inertial force causes $z$-operation in the $(x, y, z)$-coordinate system. Thus, in analyzing the inertial-force change, it is useful to utilize the $(x, y, z)$-coordinate system. The non-perturbation term, $H_1$, is described in another rotating frame by using a unitary operator

$$U_2(t) = \exp\left(-i\frac{\omega' t}{2}\sigma_z\right), \tag{9}$$

as $H_2 \equiv U_2^\dagger(t) H_1 U_2(t) - iU_2^\dagger(t)\, d/dt\, U_2(t) = (\omega_1 - \omega'/2)\sigma_z - \omega\sigma_y/2$. The eigenvectors can be described under the following condition, $|\omega| \gg |\omega' - 2\omega_1|$, as,

$$|+'\rangle_E \equiv \frac{1}{\sqrt{2(1+\delta)}}\left(i(1+\delta)|1\rangle_E + |-1\rangle_E\right), \tag{10a}$$

$$|-'\rangle_E \equiv \frac{1}{\sqrt{2(1+\delta)}}\left(|1\rangle_E + i(1+\delta)|-1\rangle_E\right), \tag{10b}$$

where $\delta \equiv 2(\omega_1 - \omega'/2)/\omega$. Under the condition $\delta \sim 0$, the diagonal Hamiltonian, $H_2$, is approximated as

$$H_2 \sim -\frac{\omega}{2}\sigma_y. \tag{11}$$

We next consider the perturbation term (Eq. (7b)) in the $(x, y, z)$-coordinate system. According to Eqs (8-9), (11), the perturbation term is described in the interaction picture of $H_2$ as

$$V_I = \exp(iH_2 t)U_2^\dagger(t)V_1(t)U_2(t)\exp(-iH_2 t) = A_\parallel \sigma_z I_z. \tag{12}$$

According to Eqs. (8-9), (11-12), the time evolution in the laboratory frame is

$$U(t) = \exp\left(-i\frac{\omega' t}{2}\sigma_z\right)\exp\left(-iA_\parallel t\, \sigma_z I_z\right). \tag{13}$$

### III. PHASE GATING OPERATION

We next consider the cyclic evolution of the electron spin. After one cyclic operation, the spin acquires a global phase, $U(\tau)|\psi(0)>_E|\uparrow(\downarrow)>_N = \exp(-i\Omega_{\uparrow(\downarrow)}(\tau))|\psi(0)>_E|\uparrow(\downarrow)>_N$, whose phase factor is determined to be

$$-i\Omega_{\uparrow(\downarrow)}(\tau) = \int_0^\tau dt \langle\uparrow(\downarrow)|\langle\psi(0)|U^\dagger(t)\left(\frac{d}{dt}U(t)\right)|\psi(0)\rangle_E|\uparrow(\downarrow)\rangle_N. \quad (14)$$

This global phase depends on the time evolution operator, $U(t)$, and on the input state of the electron spin. According to Eq. (7), the inertial force does not interact with nuclear spins; thus, the nuclear gate operation requires that the input state of electron spins depends on nuclear spin states.

In our gate sequence, the spin preparation is performed under static electric/magnetic fields, $E_0$, $B_0$. During the phase gating operation, we set the static field to zero and apply oscillating electric/magnetic fields. This switching process between static and oscillating fields is done suddenly; thus, the spin state is not changed in the process. Here, the input state of the electron spin, $|\psi(0)>_E$, corresponds to the eigenstate of the static-field Hamiltonian (Appendix B)

$$|1'\rangle_E|\uparrow(\downarrow)\rangle_N = \left(\cos\left(\frac{1}{2}\theta_{\uparrow(\downarrow)}\right)|1\rangle_E + \sin\left(\frac{1}{2}\theta_{\uparrow(\downarrow)}\right)|-1\rangle_E\right)|\uparrow(\downarrow)\rangle_N, \quad (15a)$$

$$\cos\left(\frac{1}{2}\theta_{\uparrow(\downarrow)}\right) = \frac{1}{C_{\uparrow(\downarrow)}}\left(\gamma_e B_0 + (-)\frac{A_\parallel}{2} + \sqrt{\left(\gamma_e B_0 + (-)\frac{A_\parallel}{2}\right)^2 + \left(d_\perp E_0\right)^2}\right), \quad (15b)$$

$$\sin\left(\frac{1}{2}\theta_{\uparrow(\downarrow)}\right) = \frac{d_\perp E_0}{C_{\uparrow(\downarrow)}}, \quad (15c)$$

where $C_{\uparrow(\downarrow)}$ is a normalized constant. The input state of the nuclear spin is prepared by magnetic-resonance controls as $1\sqrt{2}|1'>_E (|\uparrow>_N + |\downarrow>_N)$. According to Eq. (15), we found that using the large magnetic field is necessary to suppress decoherence under the static field because the electric-field effect induces decoherence of the nuclear spin via the hyperfine interaction. In large magnetic fields, the coherence time can exceed 1 millisecond (Appendix C), which is almost equal to coherence time of the nuclear spin limited by the relaxation time of the NV electron spin at RT ($T_{1e}$ ~5 ms) [21].

Under oscillating fields, the electron spin performs cyclic evolution, following which the electron spin incurs phase shifts depending on the nuclear spin states, $\Omega_{\uparrow(\downarrow)}$. From Eqs. (14-15), the phase factor, $\Omega_{\uparrow(\downarrow)}$, depends on the expected value $<\uparrow(\downarrow)| <1'|\sigma_z|1'>_E |\uparrow(\downarrow)>_N = \cos\theta_{\uparrow(\downarrow)}$. This is approximated under large magnetic fields, $|\gamma_e B_0| \gg |A_\parallel|, |d_\perp E_0|$, as

$$\cos\theta_{\uparrow(\downarrow)} \sim 1 - \frac{1}{2}\left(\frac{d_\perp E_0}{\gamma_e B_0 + (-)A_\parallel/2}\right)^2, \quad (16)$$

where we take lowest-order terms. The gate speed is defined as $\Delta\Omega \equiv d/dt\,(\Omega_\uparrow - \Omega_\downarrow)$, which is calculated as (Appendix D)

$$\Delta\Omega \sim \frac{\omega'}{2}\left(\frac{d_\perp E_0}{\gamma_e B_0}\right)^2 \frac{A_\parallel}{\gamma_e B_0} + A_\parallel. \quad (17)$$

Here, the nuclear spin phase is detected in a rotating frame whose angular velocity corresponds to energy splitting of the nuclear spin under the static field. Thus, the second term of Eq. (17) is canceled out. This result denotes that the gate speed is proportional to the frequency $\omega'$. When, for example, the rotational frequency of the electric field is set to $\omega'/2\pi = 1.0$ GHz, the gate time is about 165 ns (Fig. 2); this is almost equal to the theoretical limit of nuclear phase-gates by electron spin transitions [1,2]. Moreover, we can apply oscillating electric fields with frequencies in the terahertz as demonstrated in the experiments in [22]; thus, we estimate that the theoretical limit of our gate time is in principle much shorter than 100 ns.

## IV. DECOHERENCE

Geometric phase shifts are, in general, very sensitive to random fluctuations about the path. The fluctuations are mainly caused by noisy magnetic fields of nuclear spin baths [23,24]. The noise field, $\delta B$, can be described by adding it to the static magnetic field, $B_0+\delta B$. The effective noise-field amplitude to the nuclear spin, can be defined from the nuclear phase shifts, $\Delta\Omega(B_0)$, in Eq. (17) as $\gamma_n \delta B_N \equiv |\Delta\Omega(B_0+\delta B) - \Delta\Omega(B_0)| \sim 3\Delta\Omega(B_0) \delta B/B_0$, where we assume that the noise amplitude is sufficiently small, $|\delta B/B_0| \ll 1$. In addition, the random fluctuation can be described by a correlation function of random classical fields, $f(t)$ [25]. The field average is zero, $\langle f(t)\rangle = 0$, and the auto-correlation is $\langle f(t) f(0)\rangle = \exp(-t/\tau_c)$, where $\tau_c$ is the correlation time of the nuclear spin bath. The nuclear noise Hamiltonian is then represented as, $H^N_{\text{noise}} = \gamma_n \delta B_N f(t) I_z$.

Decoherence rate can be estimated by second-order calculations of the von Neumann equation. We set the input state of the nuclear spin as $\rho_N(0) = |+\rangle\langle+|_N$, where $|+\rangle_N$ is a superposition state, $|+\rangle_N = 1/\sqrt{2}\,(|\uparrow\rangle_N + |\downarrow\rangle_N)$. The ensemble-averaged density matrix $\bar{\rho}_N(t)$ shows nuclear spin coherence, which is approximated as $\langle +|\bar{\rho}_N(t)|+\rangle_N \sim \exp(-\xi(t))$, where $\xi(t) \sim (\gamma_n \delta B_N t/2)^2$ under slow fluctuations of the nuclear spin bath, $\tau_c \gg t$ [25]. The coherence time without echo is

$$\frac{1}{T^*_{2N}} = \frac{3\omega'}{4}\left(\frac{d_\perp E_0}{\gamma_e B_0}\right)^2 \frac{A_\parallel}{\gamma_e B_0}\frac{\delta B}{B_0}. \tag{18}$$

Here, gate errors by decoherence can be estimated as $\varepsilon_{\text{dec}} = 1 - \text{fidelity}$; fidelity is defined by $tr(\rho_N(t)\rho_N'(t))$ [21], where $\rho_N(t)$ is the ideal density matrix, and $\rho'_N(t)$ is the density matrix with errors. The fidelity is calculated as $1/2\,(1+\exp(-(t/T^*_{2N})^2))$. According to Eq. (17), the gate error of the nuclear phase-gate can be approximated as

$$\varepsilon_{\text{dec}}(B_0) \sim \left(\frac{3\pi}{2\sqrt{2}}\frac{\delta B}{B_0}\right)^2. \tag{19}$$

where we take lowest-order terms. In the simulation shown in Fig. 3(a), we assume a nuclear spin bath composed of $^{13}$C at a concentration of 0.03%, and the noise amplitude is $\gamma_e \, \delta B / 2\pi = 0.02$ MHz [23]. Based on this, it is seen that large static magnetic fields allow for robust spin control.

## V. SYSTEMATIC ERRORS

Robustness against systematic errors is of practical importance in quantum information processing [26,27]. Here, we assume that systematic errors can be represented as an unwanted shift of the static electric/magnetic fields, $E_0 \to E_0 + \Delta E$, $B_0 \to B_0 + \Delta B$. In the proposed method, we focus on nuclear spins, and the gate error can be also estimated as $\varepsilon_{\text{sys}} = 1 - \text{fidelity}$. The systematic error causes unwanted unitary transformations, but does not cause decoherence; thus, $\rho_N$ and $\rho'_N$ are calculated as pure states. The systematic error of the nuclear phase-gate can be described as

$$\varepsilon_{\text{sys}}(\Delta B, \Delta E) \sim 4\pi^2 \left( -\frac{3\Delta B}{4B_0} + \frac{\Delta E}{2E_0} \right)^2, \qquad (20)$$

where we assume that the unwanted shifts are sufficiently small, $|\Delta E/E_0|, |\Delta B/B_0| \ll 1$, and take lowest-order terms. Based on this, it is seen that large static electric/magnetic fields allow for robust spin control (Fig. 3(b)).

## VI. CONDITIONAL GATE

Conditional phase-gates are necessary for quantum computation. Here, we show $^{15}$N nuclear phase-gates controlled by nuclear spins of a third-nearest-neighbor carbon-13 ($^{13}$C). The hyperfine constant of the $^{13}$C ($^{15}$N) nucleus is described as $A_{\parallel}^C$ ($A_{\parallel}^N$), where $A_{\parallel}^C /2\pi = 14$ MHz [23]. This operation requires phase shifts that differ depending on the $^{13}$C nuclear spin state, $\Delta\Omega_{C=\pm}$. To calculate these shifts, we add hyperfine coupling of the $^{13}$C nucleus to the effective magnetic field as follows: $\gamma_e B_0 \rightarrow \gamma_e B_0 \pm A_\parallel^C /2$. Here, we assume that the static magnetic field satisfies $| A_\parallel^N/ (2\gamma_e B_0 \pm A_\parallel^C)| \ll 1$, and from Eq. (17), the relative phase shift is

$$\Delta\Omega_{C=+} - \Delta\Omega_{C=-} \sim -\frac{\omega'}{2} \frac{A_\parallel^N (d_\perp E_0)^2}{(\gamma_e B_0)^3} \left( \sum_{k=1} 2k(2k+1) \left( \frac{A_\parallel^C}{2\gamma_e B_0} \right)^{2k-1} \right). \qquad (21)$$

The conditional gate corresponds to $(\Delta\Omega_{C=+} - \Delta\Omega_{C=-}) t = \pi$. When we set the respective parameters to $\omega'/2\pi = 1.0$ GHz, $d_\perp E_0 / 2\pi = 4.0$ MHz, and $\gamma_e B_0 / 2\pi = 40$ MHz, the conditional gate time is about 1 μs.

## VII. CONCLUSION

We have proposed a phase-gate of single nuclear spins controlled by fast/slow rotating fields. The nuclear gate time is, in principle, much shorter than the previously reported, which is limited by the hyperfine constant [1]. We showed the robustness of the proposed method against decoherence and systematic errors. Multi-nuclear operation was also confirmed. It should be noted that our methods are not limited only to NV-spin systems, and are applicable to many quantum systems such as ion trap, spins, and superconducting qubits. Our result is a significant step for outstanding operability of long-lived quantum memories.

## ACKNOWLEDGMENTS

We are grateful to Yasushi Kondo and Mikio Nakahara for useful discussions. The authors gratefully acknowledge the financial support of SCOPE, Grant-in-Aid for JSPS Fellows Grant Number 13J04142, CREST and KAKENHI (No. 15H05868, 25220601).

## APPENDIX A: THRESHOLD OF THE ROTATING-FIELD AMPLITUDE

In the proposed gate operation, the oscillation of electric/magnetic fields plays a key role. If the amplitude $\omega_1$ is too small, the oscillating-field effect is suppressed by the orthogonal static magnetic field. This is confirmed from snap-shot Hamiltonian of the electron spin,

$$H_{snap} = \omega_1 \sigma_x + \frac{\omega_0}{2}\sigma_z, \tag{A1}$$

where the parameters $\varphi$, $\omega t$ of Eq. (3) are set to $\varphi = 0$ and $\omega t = \pi/2$, and $\omega_0$ corresponds to the hyperfine constant $A_\parallel$. Under small electric/magnetic-field conditions, $|\omega_1/\omega_0| \ll 1$, the eigenstates are

$$|\psi_1\rangle_E = \frac{1}{\sqrt{\omega_1^2 + \omega_0^2}}\left(\omega_0|1\rangle_E + \omega_1|-1\rangle_E\right), \tag{A2a}$$

$$|\psi_{-1}\rangle_E = \frac{1}{\sqrt{\omega_1^2 + \omega_0^2}}\left(\omega_1|1\rangle_E - \omega_0|-1\rangle_E\right). \tag{A2b}$$

If the oscillating-field effect is completely suppressed, these eigenstates are not changed from that of non-oscillating field, $\omega_1 = 0$. The approximation of Eq. (2) denotes that, under the following condition, the oscillating-field effect should be calculated without neglecting hyperfine off-diagonal terms

$$\left|\langle -1|\psi_1\rangle_E\right|^2 \sim \left|\langle 1|\psi_{-1}\rangle_E\right|^2 \sim \left(\frac{\omega_1}{\omega_0}\right)^2 \leq 10^{-7}. \tag{A3}$$

In this paper, the amplitude of oscillating fields satisfies the condition, $(\omega_1/\omega_0)^2 \gg 10^{-7}$.

## APPENDIX B: STATIC-FIELD HAMILTOIAN

The static-field Hamiltonian is, from Eq. (2), described as

$$H_{sta} = \gamma_e B_0 \sigma_z + d_\perp E_0 \sigma_x + A_\parallel \sigma_z I_z, \tag{B1}$$

where $B_0$ and $E_0$ are static magnetic and electric fields, respectively. The diagonal Hamiltonian is

$$H_{sta} = \sqrt{\left(\gamma_e B_0 + \frac{A_\parallel}{2}\right)^2 + (d_\perp E_0)^2} \, |1'\rangle\langle1'|_E |\uparrow\rangle\langle\uparrow|_N + \sqrt{\left(\gamma_e B_0 - \frac{A_\parallel}{2}\right)^2 + (d_\perp E_0)^2} \, |1'\rangle\langle1'|_E |\downarrow\rangle\langle\downarrow|_N$$
$$- \sqrt{\left(\gamma_e B_0 + \frac{A_\parallel}{2}\right)^2 + (d_\perp E_0)^2} \, |-1'\rangle\langle-1'|_E |\uparrow\rangle\langle\uparrow|_N - \sqrt{\left(\gamma_e B_0 - \frac{A_\parallel}{2}\right)^2 + (d_\perp E_0)^2} \, |-1'\rangle\langle-1'|_E |\downarrow\rangle\langle\downarrow|_N \quad \text{(B2)}$$

and each electron eigenstate is

$$|1'\rangle_E |\uparrow(\downarrow)\rangle_N = \frac{1}{C_{\uparrow(\downarrow)}} \left( \left[ \gamma_e B_0 + (-)\frac{A_\parallel}{2} + \sqrt{\left(\gamma_e B_0 + (-)\frac{A_\parallel}{2}\right)^2 + (d_\perp E_0)^2} \right] |1\rangle_E + d_\perp E_0 |-1\rangle_E \right) |\uparrow(\downarrow)\rangle_N, \quad \text{(B3a)}$$

$$|-1'\rangle_E |\uparrow(\downarrow)\rangle_N = \frac{1}{C_{\uparrow(\downarrow)}} \left( d_\perp E_0 |1\rangle_E - \left[ \gamma_e B_0 + (-)\frac{A_\parallel}{2} + \sqrt{\left(\gamma_e B_0 + (-)\frac{A_\parallel}{2}\right)^2 + (d_\perp E_0)^2} \right] |-1\rangle_E \right) |\uparrow(\downarrow)\rangle_N, \quad \text{(B3b)}$$

where $C_{\uparrow(\downarrow)}$ is a normalized constant. The electron eigenstates depend on the nuclear spin states. Here, each factor can be approximated under large magnetic fields, $|\gamma_e B_0| \gg |d_\perp E_0|, |A_\parallel|$, as

$$\sqrt{\left(\gamma_e B_0 \pm \frac{A_\parallel}{2}\right)^2 + (d_\perp E_0)^2} \sim \gamma_e B_0 \left(1 + \frac{1}{2}\left(\frac{d_\perp E_0}{\gamma_e B_0}\right)^2\right) \pm \frac{A_\parallel}{2}\left(1 - \frac{1}{2}\left(\frac{d_\perp E_0}{\gamma_e B_0}\right)^2\right), \quad \text{(B4)}$$

where we take lowest-order terms. Thus, the static-field Hamiltonian is re-written as

$$H_{sta} \sim \gamma_e B_0 \left(1 + \frac{1}{2}\left(\frac{d_\perp E_0}{\gamma_e B_0}\right)^2\right) S'_Z + A_\parallel \left(1 - \frac{1}{2}\left(\frac{d_\perp E_0}{\gamma_e B_0}\right)^2\right) S'_Z I_Z, \quad \text{(B5)}$$

where $S_z'$ is defined as $S_z' \equiv |1'\rangle\langle1'|_E - |-1'\rangle\langle-1'|_E$.

## APPENDIX C: DECOHERENCE UNDER STATIC FIELDS

Noisy magnetic fields cause decoherence. Here, we assume that the noisy magnetic fields are sufficiently small, $|\delta B/B_0| \ll 1$. By adding a noisy magnetic field term as $B_0 + \delta B$, Eq. (B4) becomes

$$\sqrt{\left(\gamma_e B_0 \pm \frac{A_\parallel}{2} + \gamma_e \delta B\right)^2 + \left(d_\perp E_0\right)^2} \,. \tag{C1}$$

Under large static magnetic fields, this is approximated as

$$\sim \sqrt{\left(\gamma_e B_0 \pm \frac{A_\parallel}{2}\right)^2 + \left(d_\perp E_0\right)^2} \left(1 + \gamma_e \delta B \frac{\gamma_e B_0 \pm A_\parallel/2}{\left(\gamma_e B_0 \pm A_\parallel/2\right)^2 + \left(d_\perp E_0\right)^2}\right). \tag{C2}$$

Thus, the noise terms are separated from the non-noise terms. Using the following approximations

$$\sqrt{\left(\gamma_e B_0 \pm \frac{A_\parallel}{2}\right)^2 + \left(d_\perp E_0\right)^2} \sim \left(\gamma_e B_0 \pm \frac{A_\parallel}{2}\right)\left(1 + \frac{1}{2}\left(\frac{d_\perp E_0}{\gamma_e B_0 \pm A_\parallel/2}\right)^2\right), \tag{C3a}$$

$$\frac{\gamma_e B_0 \pm A_\parallel/2}{\left(\gamma_e B_0 \pm A_\parallel/2\right)^2 + \left(d_\perp E_0\right)^2} \sim \frac{1}{\gamma_e B_0 \pm A_\parallel/2}\left(1 - \left(\frac{d_\perp E_0}{\gamma_e B_0 \pm A_\parallel/2}\right)^2\right), \tag{C3b}$$

the noise terms of Eq. (C2) can be re-written as

$$\gamma_e \delta B \left(1 - \frac{1}{2}\left(\frac{d_\perp E_0}{\gamma_e B_0}\right)^2 \left(1 \mp \frac{A_\parallel}{\gamma_e B_0}\right)\right). \tag{C4}$$

where we take lowest-order terms. The noise static-field Hamiltonian is shown as

$$H_{sta}^{noise} \sim \gamma_e \delta B \left(1 - \frac{1}{2}\left(\frac{d_\perp E_0}{\gamma_e B_0}\right)^2\right) S_z' + \gamma_e \delta B \left(\frac{d_\perp E}{\gamma_e B_0}\right)^2 \frac{A_\parallel}{\gamma_e B_0} S_z' I_z \,. \tag{C5}$$

In the proposed method, the electron spin is in its eigenstates under static electric/magnetic fields, and we focus on nuclear spin decoherence. From Eq. (C5), the effective noise Hamiltonian of the nuclear spin can be described as,

$$V_{noise}^N = b f(t) I_z, \tag{C6}$$

where the noise amplitude $b$ is,

$$b = \gamma_e \delta B \left(\frac{d_\perp E_0}{\gamma_e B_0}\right)^2 \frac{A_\parallel}{\gamma_e B_0}, \tag{C7}$$

where $f(t)$ is the correlation function described in the section 4.

The initial state of the nuclear spin is set as $\rho_N = |+\rangle\langle+|_N$, where $|+\rangle_N = 1/\sqrt{2}\,(|\uparrow\rangle_N + |\downarrow\rangle_N)$. The time evolution can be estimated by second-order calculations of the von Neumann equation. From the similar calculation of Eq. (18), coherence time of the nuclear spin is estimated as,

$$\frac{1}{T_{2N}^*} \sim \frac{b}{2} = \frac{\gamma_e \delta B}{2}\left(\frac{d_\perp E_0}{\gamma_e B_0}\right)^2 \frac{A_\parallel}{\gamma_e B_0}. \tag{C8}$$

In the simulation shown in Fig. 4, we assume the nuclear spin bath composed of $^{13}$C at a concentration of 0.03% and the noise amplitude of $\gamma_e \delta B / 2\pi = 0.02$ MHz. The coherence time is longer than 1 ms, which is almost equal to the nuclear coherence time limited by the relaxation time of the NV electron spin at RT ($T_{1e} \sim 5$ ms).

## APPENDIX D: PHASE SHIFT OF THE NUCLEAR SPIN

The gate speed, $\Delta\Omega \equiv d/dt\,(\Omega_\uparrow - \Omega_\downarrow)$, is, from Eq. (13-16), calculated as

$$\Delta\Omega = A_\parallel - \frac{d_\perp E_0}{2}\left(\frac{d_\perp E_0}{\gamma_e B_0 + A_\parallel/2} - \frac{d_\perp E_0}{\gamma_e B_0 - A_\parallel/2}\right) - \frac{\omega'}{4}\left(\left(\frac{d_\perp E_0}{\gamma_e B_0 + A_\parallel/2}\right)^2 - \left(\frac{d_\perp E_0}{\gamma_e B_0 - A_\parallel/2}\right)^2\right). \tag{D1}$$

Under large magnetic fields, $|\gamma_e B_0| \gg |d_\perp E_0|, |A_\parallel|$, the second terms are approximated as

$$\frac{d_\perp E_0}{2}\left(\frac{d_\perp E_0}{\gamma_e B_0 + A_\parallel/2} - \frac{d_\perp E_0}{\gamma_e B_0 - A_\parallel/2}\right) \sim -\frac{A_\parallel}{2}\left(\frac{d_\perp E_0}{\gamma_e B_0}\right)^2. \tag{D2}$$

This term is much smaller than the first term, and we thus neglect it. The third terms are also approximated as

$$\frac{\omega'}{4}\left(\left(\frac{d_\perp E_0}{\gamma_e B_0 + A_\parallel/2}\right)^2 - \left(\frac{d_\perp E_0}{\gamma_e B_0 - A_\parallel/2}\right)^2\right) \sim -\frac{\omega'}{2}\left(\frac{d_\perp E_0}{\gamma_e B_0}\right)^2 \frac{A_\parallel}{\gamma_e B_0}. \tag{D3}$$

Thus, the gate speed can be described as

$$\Delta\Omega \sim \frac{\omega'}{2}\left(\frac{d_\perp E_0}{\gamma_e B_0}\right)^2 \frac{A_\parallel}{\gamma_e B_0} + A_\parallel. \tag{D4}$$

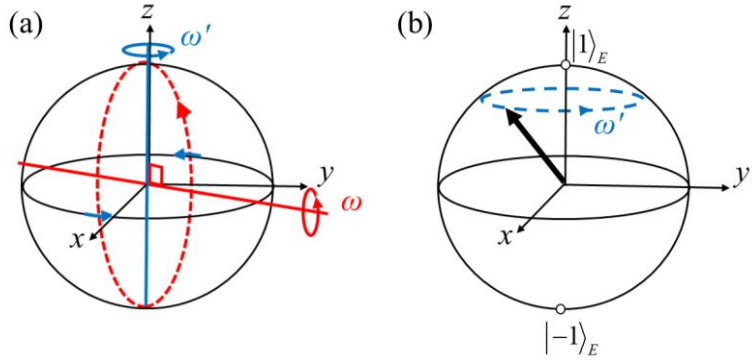

FIG. 1 T. Shimo-Oka et al.

FIG. 1(a) The inertial force is generated by the rotating field in the laboratory frame (red arrow). Gate operations are performed by adiabatic rotations of the inertial force (blue arrow). (b) The adiabatic rotation of the inertial force leads to the electron spin cyclic-evolution (blue dash arrow) in the NV electron Bloch square.

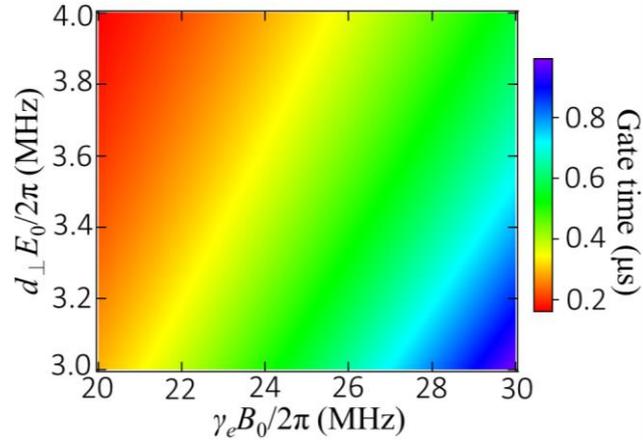

FIG. 2 T. Shimo-Oka et al.

FIG. 2. Gate time of the proposed nuclear spin phase-gate. $\gamma_e B_0$ and $d_\perp E_0$ correspond to the amplitude of the static magnetic and electric fields, respectively. The frequency of the slow electric-field rotation is $\omega' / 2\pi = 1.0$ GHz.

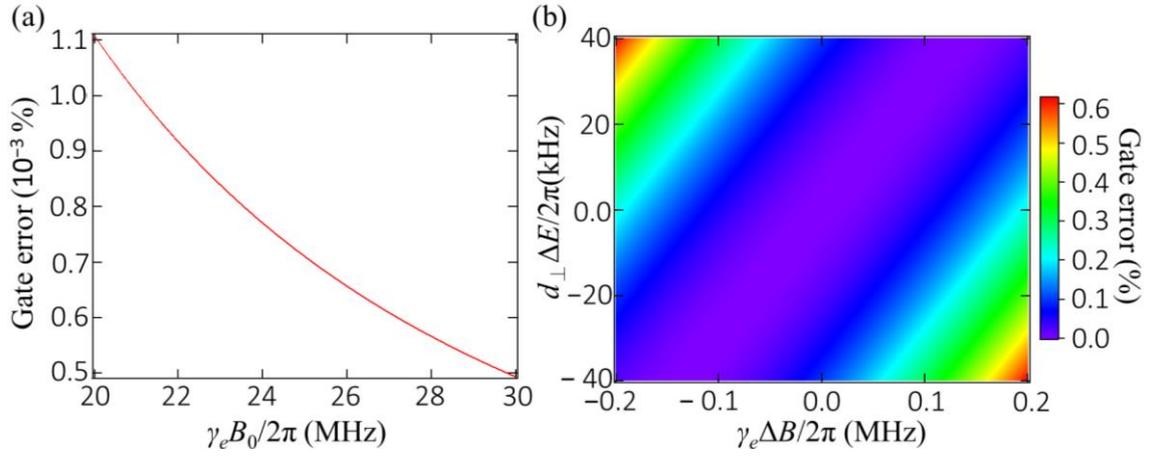

FIG. 3. T. Shimo-Oka et al.

FIG. 3. Robustness against decoherence and systematic errors. (a) Stochastic errors by decoherence: $\gamma_e B_0$ correspond to the amplitude of static magnetic fields. The noise amplitude is $\gamma_e \delta B / 2\pi = 0.02$ MHz. (b) Systematic error: $\Delta B$, $\Delta E$ denote unwanted shifts of the static magnetic and electric fields, respectively. The amplitude of the static electric/ magnetic fields are, $d_\perp E_0 / 2\pi = 4.0$ MHz, $\gamma_e B_0 / 2\pi = 20$ MHz, respectively.

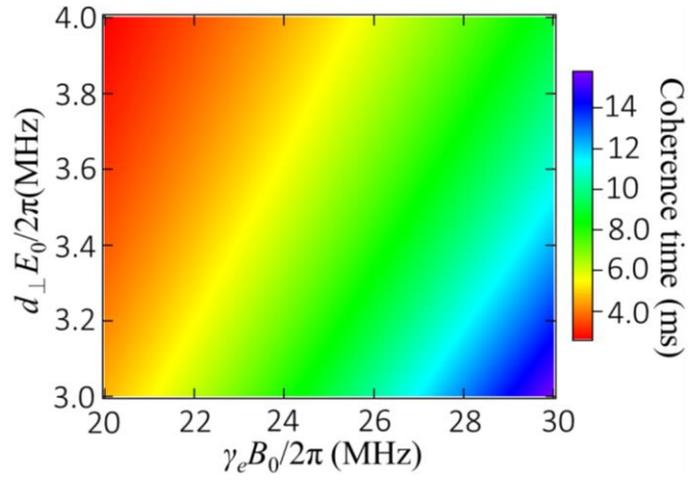

FIG. 4. T. Shimo-Oka et al.

FIG. 4. Coherence time under static electric/magnetic fields: $\gamma_e B_0$, $d_\perp E_0$ correspond to the amplitude of the static magnetic and electric fields, respectively. The noise amplitude is $\gamma_e \delta B / 2\pi = 0.02$ MHz.